\documentclass{iau} 
\usepackage{graphicx}

\title[Phoenix dT] 
{A VLT/FORS2 spectroscopic survey of individual stars in a transforming dwarf galaxy}

\author[Battaglia et al.]   
{G.Battaglia$^{1,2}$
 \and N.Kacharov$^3$ \and M.Rejkuba$^4$}

\affiliation{$^1$Instituto de Astrof\'{i}sica de Canarias, calle V\'{i}a L\'{a}ctea s/n, E38205 - La Laguna (Tenerife), Spain \\[\affilskip]
$^2$Universidad de la Laguna, Dpto. Astrofisica, 38206 La Laguna, Tenerife, Spain\\ email: {\tt gbattaglia@iac.es} \\[\affilskip]
  $^3$ Max Planck Institut f\"{u}r Astronomie, K\"{o}nigstuhl 17, 69117 Heidelberg, Germany \\ [\affilskip]
  $^4$ European Southern Observatory, Karl-Schwarzschild-Str. 2, 85748 Garching bei M\"{u}nchen, Germany}

\pubyear{2016}
\volume{321}  
\jname{Formation and evolution of galaxy outskirts}
\editors{A. Gil de Paz, J. Lee, \& J. H. Knapen, eds.}
\begin{document}

\maketitle

\begin{abstract}
Understanding the properties of dwarf galaxies is important not only to put them in their proper cosmological context, 
but also to understand the formation and evolution of the most common type of galaxies. Dwarf galaxies are divided into two main classes, dwarf irregulars (dIrrs) and dwarf spheroidals (dSphs), which differ from each other mainly because the former are gas-rich objects currently forming stars, while the latter are gas-deficient with no on-going star formation. Transition types (dT) are thought to represent dIs in the process of losing their gas, and can 
therefore shed light into the possible process of dwarf irregulars (dIrrs) becoming gas-deficient, passively evolving 
galaxies. Here we present preliminary results from 
our wide-area VLT/FORS2 MXU spectroscopic survey of the Phoenix dT, from which we obtained line-of-sight velocities 
and metallicities from the nIR Ca~II triplet lines for a large sample of individual Red Giant Branch stars. 
\end{abstract}

\firstsection 
\section{Introduction}
As in other galaxy groups, also the dwarf galaxies inhabiting the Local Group 
can be broadly divided in two categories, late- and early-types, 
on the basis of the presence or absence of both gas and current star formation, 
respectively.  It is still a question whether these morphological 
types share similar ancestors and owe their distinct properties at redshift
 zero to a different evolutionary path, or whether their ``destiny'' as 
a late- or early-type system was somewhat imprinted in the conditions at 
their formation. The fact that Local Group 
late- and early-type dwarfs share similar 
scaling relations (see \cite[Tolstoy, Hill \& Tosi 2009]{2009ARA&A..47..371T} for a recent review), 
appear to follow the same stellar mass-metallicity relation 
(\cite[Kirby et al. 2013]{2013ApJ...779..102K}) but inhabit clearly different environments - 
with late-types found in isolation and early-types in general satellites 
of the large Local Group spirals - hints to environmental effects being 
relevant for the evolution of these galaxies. 

On the basis of lifetime star formation histories from very deep 
colour-magnitude-diagrams, \cite[Gallart et al.(2015)]{2015ApJ...811L..18G} propose the existence of two 
main types of dwarf galaxies: {\it slow dwarfs}, i.e. 
those that have formed stars at a relatively low rate for the whole 
life of the system, and {\it fast dwarfs}, i.e. those that have experienced 
most of their star formation activity in the first Gyrs of evolution. 
Late-type systems map into {\it slow dwarfs}, but this is not always the 
case for early-type and {\it fast-} dwarfs. Gallart et al. propose that 
the two evolutionary paths are imprinted from the start and related to the 
environment in which the dwarfs were born. 

The spectroscopic study of the evolved stellar component of dwarf galaxies 
offers a complementary perspective, as for Local Group galaxies 
it is possible to determine metallicities and line-of-sight velocities 
for individual stars, and hence study their internal kinematics and 
metallicity properties in detail and out to large radii. 
Red giant branch stars are particularly interesting targets, 
because they probe a very large range of ages, 
from $\sim$1.5Gyr to the early star formation episodes. Being very luminous, 
they are within reach of current spectrographs on 8m-10m telescopes for galaxies 
to within $\sim$1 Mpc. 


\section{The data-set}

Here we present preliminary results from our VLT/FORS2 MXU 
spectroscopic survey of RGB stars in the Phoenix transition type (dT). This 
system is as luminous as a typical ``classical'' spheroidal, but 
found at a distance of $\sim$400kpc from the Milky Way, hence it can provide an 
interesting view of the properties of low mass dwarf galaxies that have mostly 
evolved in isolation.

We obtained VLT/FORS2 MXU spectroscopic data for the Phoenix dT in service mode (Programme 083.B-0252; PI: Battaglia). 
We used the 1028z grism in combination with the OG590+32 order separation filter (wavelength coverage 7730 - 9480\,\AA\,);
the slits were set to have a width=1'' for a final resolving power of R$\sim$2500. We obtained 11 pointings
in which we allocated slits to a total of $254$ pre-selected red giant branch (RGB; I\,$<21.5$\,mag) stars
from \cite[Battaglia et al.(2012)]{2012MNRAS.424.1113B}
that cover a total area of roughly $10' \times 20' $ across the field of the galaxy.
In addition to these data,
we use VLT/FORS2 MXU spectra for ~30 stars in a central pointing (Programme 71.B-0516; PI: Cole).

The data reduction and extraction of the spectra was performed with standard 
IRAF procedures, together with custom made IDL scripts. We extracted line-of-sight velocities (v$_{\rm l.o.s.}$) and
equivalent widths (EWs) from the Ca~II triplet (CaT) nIR lines. Line-of-sight velocities were obtained
by cross-correlating the continuum normalized spectra with a synthetic spectrum derived
from a stellar atmospheric model with similar parameters as expected for the Phx targets. The EWs were
determined by fitting a Voigt profile and transformed into [Fe/H] using the calibration from
\cite[Starkenburg et al.(2010)]{2010A&A...513A..34S}. The CaT [Fe/H] was validated
against two calibrating globular clusters; these were also used to test the accuracy of the v$_{\rm l.o.s.}$ determination,
together to Phx stars overlapping among different masks. The median error in v$_{\rm l.o.s.}$ is 8 km s$^{-1}$.
The sample of probable Phoenix members among the two observing programmes
consists of $\sim$190 stars, as indicated by the stars' position on the colour-magnitude-diagram and
radial velocity.

\section{Results}

We obtain a systemic l.o.s. velocity equal to -21.2$\pm$1.0 km s$^{-1}$ and a dispersion of 9.3$\pm$0.7 km s$^{-1}$.
The systemic velocity is consistent with the determination of -13$\pm$9 km s$^{-1}$ obtained by \cite[Irwin \& Tolstoy (2002)]{2002MNRAS.336..643I} from a
much smaller sample of 7 stars and in excellent agreement with the velocity of the HI cloud (-23 km s$^{-1}$) proposed by
\cite[St-Germain et al.(1999)]{1999AJ....118.1235S} as physically associated to Phoenix.

It is well-known that dwarf galaxies inhabiting dense environments
such as within the virial radii of the Milky Way or M31
are devoid of gas, except for relatively massive systems such as e.g. for the Magellanic Clouds or IC~10.
The fact that such a small system as Phoenix has been able to
form stars until almost present day (\cite[Hidalgo et al. 2009]{hidalgo+2009})
and still contains gas, together with its large distance and velocity direction approaching the Milky Way
(v$_{\rm GSR}= -108.6 \pm 1.0$km s$^{-1}$),
strongly suggests that Phoenix is still to enter the Milky Way virial radius for the first time and has evolved
undisturbed from interactions with large galaxies.

\begin{figure}[b]
\begin{center}
 \includegraphics[width=3.4in]{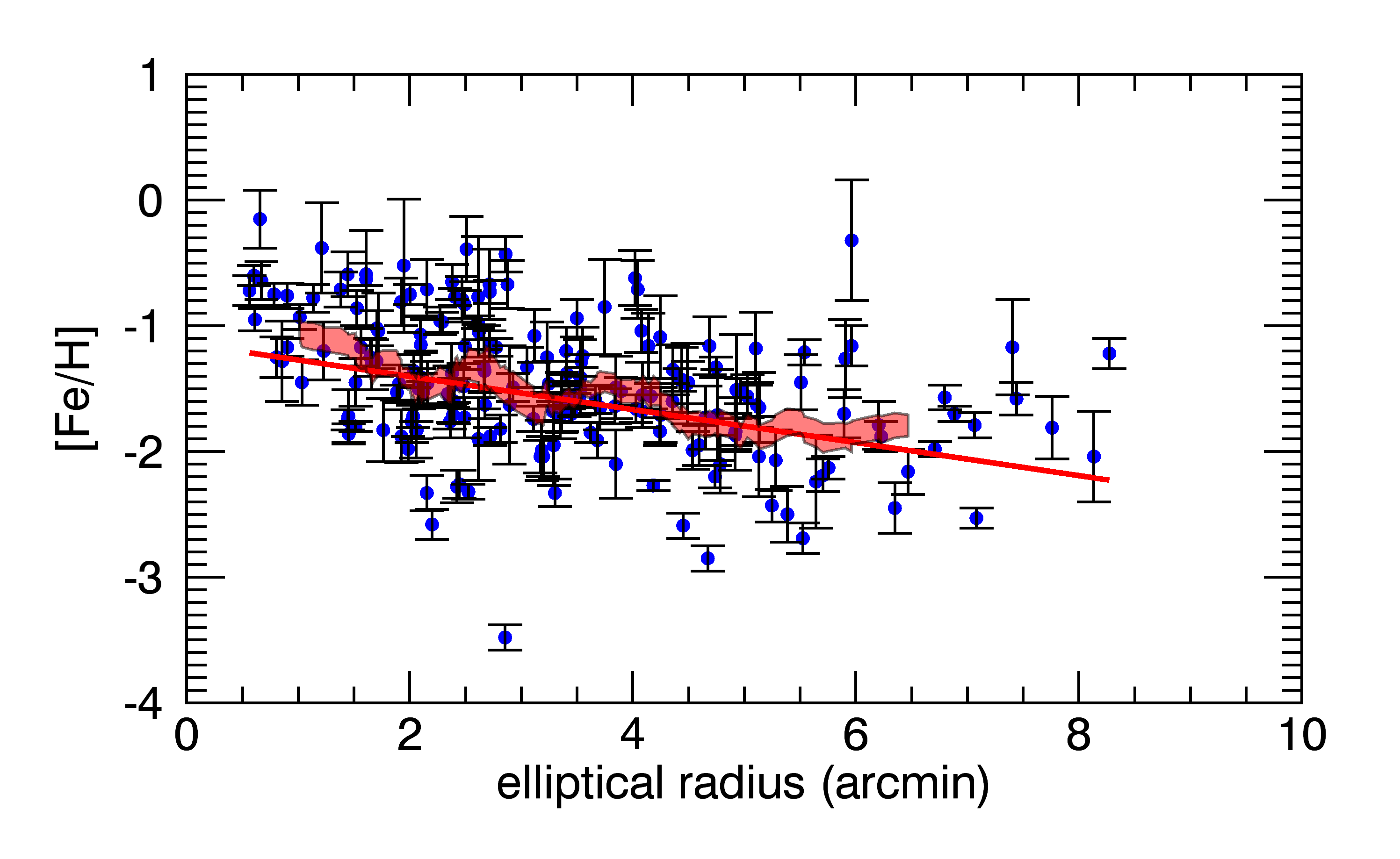} 
 \caption{Metallicity versus elliptical radius for the Phoenix probable member stars. The red lines
 show the running average and the linear fit to the metallicity properties.}
   \label{fig1}
\end{center}
\end{figure}

Interestingly, the metallicity properties of this system appear alike to those of similarly luminous
dwarf spheroidal galaxies satellites of the Milky Way, which are devoid of gas and are found in a higher
density environment.  In particular we detect a clear negative, approximately linear [Fe/H] gradient (see Fig.~\ref{fig1}), 
whose slope compares well to those of Milky Way dwarf spheroidal galaxies (e.g. \cite[Leaman et al. 2013]{2013ApJ...767..131L}).
This finding leads us to speculate that
metallicity gradients in these small dwarf galaxies are likely to be driven by internal mechanisms rather than by environmental ones,
although this conclusion needs to be put on a more secure foot by
similar spectroscopic surveys of larger samples of Local Group dwarf galaxies inhabiting different environments.

The detailed metallicity and kinematic properties of Phoenix from the VLT/FORS2 MXU sample
here presented will be discussed in Kacharov et al. (in prep.)

\section*{Acknowledgements}
We are grateful to A.Cole for useful discussions and for providing the reduced
VLT/FORS2 data from program 71.B-0516.


\end{document}